\documentclass[aps,prl,showpacs,twocolumn,superscriptaddress,longbibliography]{revtex4-1}
\usepackage[pdftex]{graphicx}
\usepackage[pdftex,bookmarks,colorlinks,breaklinks]{hyperref}  
\hypersetup{linkcolor=red,citecolor=blue,filecolor=dullmagenta,urlcolor=blue} 
\usepackage[toc,page]{appendix}
\usepackage{amsmath,color}
\usepackage{dsfont}

\usepackage{soul}

\newcommand{\derd}{{\rm d}}

\begin{document}

\setlength{\tabcolsep}{1pt}
\title{Focusing inside Disordered Media with the Generalized Wigner-Smith Operator}

\author{Philipp Ambichl}
\affiliation{Institute for Theoretical Physics,
Vienna University of Technology (TU Wien), A-1040 Vienna, Austria, EU}
\author{Andre Brandst\"otter}
\affiliation{Institute for Theoretical Physics,
Vienna University of Technology (TU Wien), A-1040 Vienna, Austria, EU}
\author{Julian B\"ohm}
\affiliation{Universit\'{e} C\^{o}te d'Azur, CNRS, LPMC, 06108 Nice, France, EU}
\author{Matthias K\"uhmayer}
\affiliation{Institute for Theoretical Physics,
Vienna University of Technology (TU Wien), A-1040 Vienna, Austria, EU}
\author{Ulrich Kuhl}
\affiliation{Universit\'{e} C\^{o}te d'Azur, CNRS, LPMC, 06108 Nice, France, EU}
\author{Stefan Rotter}
\affiliation{Institute for Theoretical Physics,
Vienna University of Technology (TU Wien), A-1040 Vienna, Austria, EU}
\begin{abstract}
We introduce a wavefront shaping protocol for focusing inside disordered media based on a generalization of the established Wigner-Smith time-delay operator. The key ingredient for our approach is the scattering (or transmission) matrix of the medium and its derivative with respect to the position of the target one aims to focus on. A specific experimental realization in the microwave regime is presented showing that the eigenstates of a corresponding operator are sorted by their focusing strength -- ranging from strongly focusing on the designated target to completely bypassing it. Our protocol works without optimization or phase-conjugation and we expect it to be particularly attractive for optical imaging in disordered media.
\end{abstract}

\maketitle
One of the most formidable challenges for imaging in complex environments is to overcome the limitations imposed by the presence of disorder. In particular, if the wave scattering induced by a disordered medium is strong enough to suppress the ballistic contribution in the imaging process entirely, seeing through this medium or focusing on a target inside of it becomes a highly non-trivial exercise -- a difficulty that is particularly evident in the field of biological and medical imaging. A promising new approach to image and focus also in the regime of multiple scattering is to exploit the information stored in a system's scattering matrix \cite{Popoff2010tm, Yu2013}. In this emerging new field of ``wavefront shaping'' \cite{Mosk2012, rott16} spectacular advances have recently been made, such as to focus light behind an opaque layer \cite{Vellekoop2010, Vellekoop2007, Katz2011, McCabe2011, Aulbach2012}, or to send and retrieve images across it \cite{Popoff2010im, Mounaix2016, Mounaix2016b, Paudel2013, Aulbach2011}. Thanks to these advances also the focusing of light \textit{inside} highly disordered media could recently be demonstrated using embedded fluorescent probes and nano-crystals \cite{Vellekoop2008, Aulbach2012b, Chaigne2014} or using digital optical phase conjugation to focus light onto a target moving inside an otherwise static environment \cite{Xu2011, Ma2014, Judkewitz2013, Zhou2014}.

Here we present a new approach for focusing inside a disordered material that has the considerable advantage of working without the requirement to implant a fluorescent body at the focus or to phase-conjugate a wave scattered at the focus position. Our technique also allows us to tune the degree of focus on a designated position inside the disorder, including the case where the target is entirely avoided by the scattered wavefront. Our starting point for achieving this goal is the time-delay operator $Q$ introduced by Eugene Wigner and Felix Smith \cite{Wigner1955, Smith1960}. Originally devised for nuclear scattering problems to deduce the time associated with a scattering event from stationary measurements of the asymptotic scattering amplitudes, this concept prominently resurfaced in mesoscopic physics \cite{Kottos2005} and very recently in attosecond physics \cite{Pazourek2015} as well as in the newly emerging community of wavefront shaping \cite{rott16, Fan2005, Rotter2011, Carpenter2015, Xiong2016, Gerardin2016}.

The Wigner-Smith time-delay operator $Q$ is constructed based on a system's scattering matrix $S$ by way of a frequency derivative, $Q = -i S^{-1} \derd S/\derd \omega$. The eigenvalues of $Q$, also called ``proper delay times'', measure the time-delay associated with the scattering by a given potential {\cite{rott16, Fyodorov1997, Brouwer1997, Savin2003}. The corresponding eigenvectors, also called ``principal modes'', are states that can be associated with this well-defined time-delay -- a property that makes them dispersion-free \cite{Fan2005} in the sense that a small variation of their input frequency does not change their spatial output profile. Moreover, principal modes have been shown to be ``particle-like'' in situations where ballistic scattering occurs \cite{Rotter2011}. The potential for applications of these states as efficient communication channels in systems with multiple in- and output ports has recently taken center stage when first experiments reported on the successful implementation of principal modes in optical multi-mode fibers \cite{Carpenter2015, Xiong2016} as well as in resonant scattering media \cite{Gerardin2016,BoehmToBePublished}.

Here we demonstrate that the concept underlying the principal modes is not at all restricted to the time-delay operator $Q$ from above; specifically, we show that in the same way as the conventional principal modes  are invariant with respect to a frequency variation, we may also create wave states that are invariant with respect to changes in the system configuration, like a local shift of a designated scatterer inside a disordered medium. While the frequency-insensitive principal modes are the eigenstates of the time-delay operator $Q = -i S^{-1} \derd S/\derd \omega$ (involving a frequency derivative), the class of states we introduce will be the eigenstates of a corresponding operator $Q_\alpha = -i S^{-1} \derd S/\derd \alpha$, where the parameter $\alpha$ stands, e.g., for the position of a movable scatterer. What is special about our new approach is that the eigenstates of $Q_\alpha$ are not only insensitive to a small change of $\alpha$, but that the associated eigenvalues indicate how strongly the corresponding conjugate quantity to $\alpha$ is affected by the scattering process. This insight allows us to maximally focus on or maximally avoid a specific target inside multiply scattering media just based on the medium's scattering matrix $S$ and its variation due to a shift of the target particle that one aims to focus on.

We start by reviewing the principal modes' remarkable property of being insensitive with respect to a change in frequency of the incident wave. The corresponding eigenvalue equation for the principal modes $\vec{u}_n$ (given as a coefficient vector in a certain basis) and for the proper delay times $\theta_n$ reads as follows,
\begin{equation}\label{eq:Q-eigenvalue}
Q \,\vec{u}_n = -i S^{-1} \frac{\derd S}{\derd \omega} \,\vec{u}_n = \theta_n \,\vec{u}_n\,,
\end{equation}
with $\omega$ being the frequency. In a waveguide system, as shown in Fig.~\ref{fig:setup}, the incident principal modes can be decomposed into waves injected from the left and right lead respectively, i.e., $\vec{u}_n = (\vec{u}_{n,L},\,\vec{u}_{n,R})^T$. For a unitary scattering matrix, $S^\dagger S= \mathds{1}$, the time-delay operator $Q$ is Hermitian with real eigenvalues $\theta_n$. Using the input-output relation $\vec{v}_n(\omega) = S(\omega) \vec{u}_n$, we can rewrite Eq.~\eqref{eq:Q-eigenvalue} for a (static) time-delay eigenstate $\vec{u}_n$ evaluated at a chosen frequency $\omega_0$
\begin{equation}\label{eq:static-output}
\left.\frac{\derd\vec{v}_n}{\derd\omega}\right|_{\omega_0} \!\!\!\!\!\! =
i \theta_n \vec{v}_n\!\left(\omega_0\right) \rightarrow
\vec{v}_n\!\left(\omega_0\!+\! \Delta\omega\right) \!\approx\!
e^{i \theta_n \Delta\omega} \, \vec{v}_n\!\left(\omega_0\right)\!.
\end{equation}
The expression on the right-hand side of Eq.~\eqref{eq:static-output} states that the output vector of a principal mode, ${\vec{v}_n = (\vec{v}_{n,L},\,\vec{v}_{n,R})^T}$, retains its original orientation when shifting the input frequency in the vicinity of $\omega_0$. This, in turn, translates into output patterns that are invariant (to first order) with respect to a frequency change (apart from a complex global factor). An aspect of the above derivation that  has so far been unexploited is the fact that the derivative in Eq.~\eqref{eq:Q-eigenvalue} does not necessarily have to be taken with respect to the frequency. In other words: the symbol $\omega$ can in principle represent any other parametric dependence of the scattering matrix and the stability property of the corresponding principal modes will still hold with respect to a variation of this new parameter. Accordingly, we shift our attention to a whole class of generalized Wigner-Smith (GWS) operators $Q_\alpha=-i S^{-1} \derd S/\derd \alpha$ in which the frequency $\omega$ is replaced by the arbitrary parameter $\alpha$. We also expect that these GWS operators may have interesting connections to earlier works where statistical properties of parametric variations of the scattering matrix have been studied \cite{lee85b, gor06c, Fyodorov2012}.

While the eigenstates of $Q_\alpha$ are invariant with respect to a small parametric shift of $\alpha$ already by construction [as in Eq.~\eqref{eq:static-output} with $\omega\to\alpha$], we still need to clarify how to interpret the corresponding eigenvalues $\theta^\alpha_n$.
Already from the dimensions it is clear that $\theta^\alpha_n$ must be associated with the conjugate variable to $\alpha$, in the same way as the delay time $\theta$ is the conjugate quantity to the frequency $\omega$.
To make this more evident we now define $C_\alpha := -i \derd/\derd\alpha$ as the corresponding conjugate operator and assume that the parameter $\alpha$ is global. It follows after a short derivation (see supplemental material \cite{Ambichl2016}),
\begin{equation}\label{eq:D-expectation-value}
\vec{u}^\dagger Q_\alpha \vec{u} =
\vec{u}^{\,\dagger} C_\alpha \vec{u} - \vec{v}^\dagger C_\alpha \vec{v}  =
\left\langle C_\alpha \right\rangle_{\rm in} - \left\langle C_\alpha \right\rangle_{\rm out} ,
\end{equation}
where $\vec{v} (\alpha) = S(\alpha) \vec{u}$ is the output vector and $\langle \ \ldots \  \rangle_\mathrm{in/out}$ denotes the expectation value evaluated in the input/output scattering state. Following Eq.~\eqref{eq:D-expectation-value}, $Q_\alpha$ is the appropriate operator to measure the shift in the conjugate variable to $\alpha$, which the wave experiences due to the scattering process -- in perfect analogy to the time-delay operator measuring a shift in time, i.e., in the conjugate quantity of the frequency $\omega$. One specific example for such a global parameter $\alpha$ could be the displacement in $y$-direction (i.e., $\alpha = y$) of the entire scattering landscape from its initial position. The conjugate variable to the position is the momentum. The eigenvalues $\theta^y_n$ of the operator $Q_y$ then measure the momentum shift (in $y$-direction) that the corresponding eigenstates experience as a result of scattering at the entire potential landscape. Solving the eigenvalue problem for the operator $Q_y$ thus provides us with an orthogonal and complete basis of eigenstates that are sorted by this momentum shift. To be more precise, the operator $Q_y$ generally measures the shift in the \textit{wavenumber} in displacement direction, but for the sake of simplicity we refer to this shift as \textit{momentum} shift $\Delta k$.

\begin{figure}
\centerline{\includegraphics[width=0.98\linewidth]{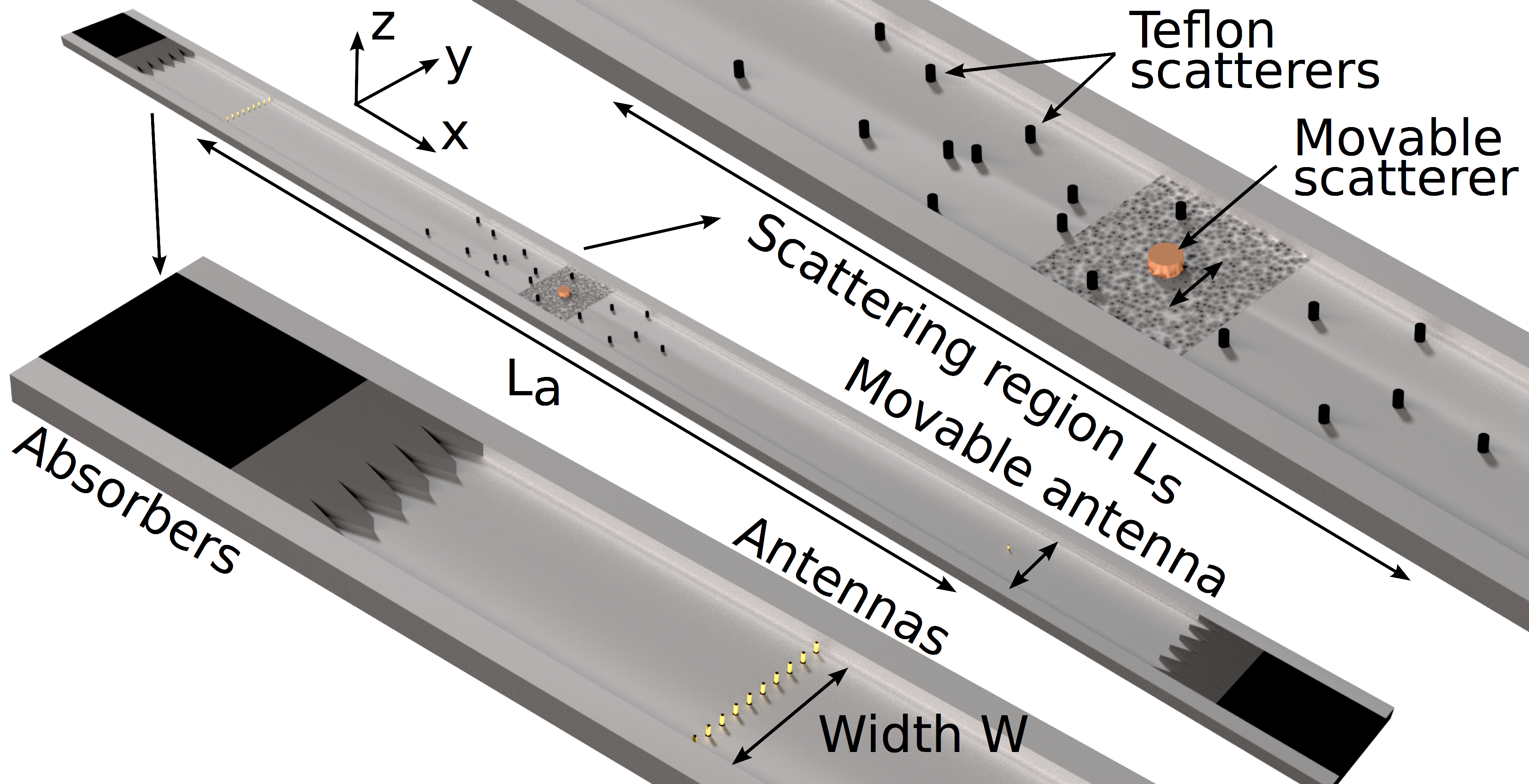}}
\caption{\label{fig:setup}
(Color online.)  Sketch of the experimental setup. The system consists of a rectangular aluminum waveguide of height $H=8$\,mm, width $W=10$\,cm, and total length $L=2.38$\,m, see middle panel. The top plate (not shown) can be removed. The wavefront is injected from the left using ten monopole antennas, see bottom panel. The scattering region displayed in the top panel has a length $L_s=60$\,cm and consists of 18 randomly placed cylindrical Teflon scatterers (black cylinders, index of refraction $n=1.44$, radius 2.55\,mm). A (re-)movable brass scatterer of radius 8.825\,mm is located in the central part of the scattering region. The placement of scatterers in this sketch matches the actual scatterer positions in the experiment. The distance between the injecting antenna array and the scanning antenna is $L_a=1.50$\,m. The grained area around the movable scatterer indicates the region shown in Fig.~\ref{fig:focus} and \ref{fig:omission}.}
\end{figure}

In the present context, we will be specifically interested in the case where $\alpha$ is the position $y_i$ of the $i$-th scattering element inside a strongly disordered medium consisting of altogether $i_{\rm tot}\geq i$ such individual elements (see, Fig.~\ref{fig:setup})}. In this case, $\alpha$ does not represent a global variable since only the $i$-th scatterer is shifted rather than the whole system. It turns out that Eq.~\eqref{eq:D-expectation-value} still remains valid for this more general case, but that the corresponding expectation values are then evaluated based on the local field amplitudes in the vicinity of the $i$-th scatterer (see derivation in \cite{Ambichl2016}). The eigenvalues $\theta^{y_i}_n$ of $Q_{y_i}$ are thus equal to the momentum shift $\Delta k^i$ that the wave experiences \textit{locally} when colliding with the $i$-th scatterer (in the presence of all other scatterers and boundaries of the medium). Note that a large momentum shift requires that the wavefront of the incident eigenstate is back-scattered significantly at the target position $x_i$, which is equivalent to focusing on the $i$-th scatterer. On the other hand, when the wave does not get scattered at the target at all (e.g., by not reaching the $i$-th scatterer), the momentum transfer will, correspondingly, be very small. In this way the GWS operator $Q_{y_i}$ provides us the means to focus on a designated scatterer inside a disordered medium or to omit this target simply by generating an eigenvector of the GWS operator corresponding to a large or small eigenvalue, respectively.

\begin{figure}
\centerline{\includegraphics[width=0.94\linewidth]{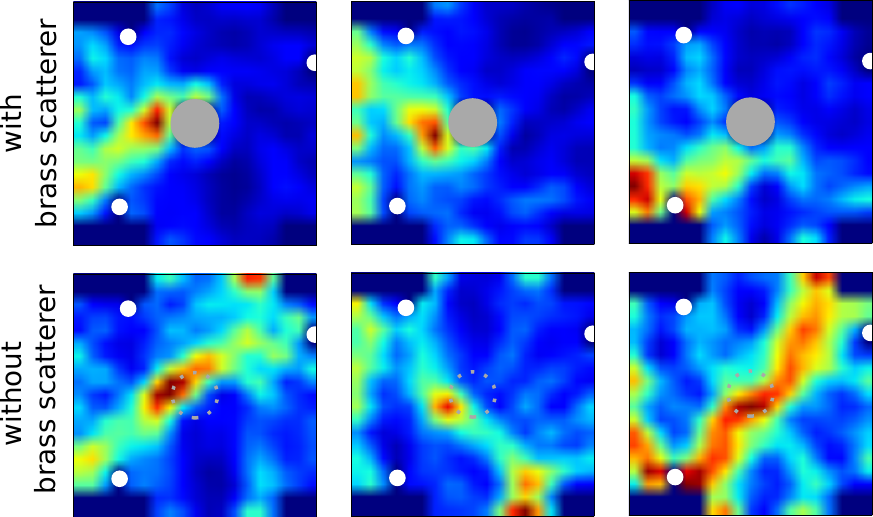}}
\caption{\label{fig:focus}
(Color online.) Measured spatial distribution of the microwave intensity (see supplemental material \cite{Ambichl2016}) inside the disordered scattering system realized in the experiment. The region shown is a zoom on the vicinity of the movable brass scatterer in the middle (as highlighted in the top panel of Fig.~\ref{fig:setup}). In the top row this brass scatterer is included and in the bottom row it is removed. In both cases we display the eigenstates of $q_{y_{\rm b}}$ with the largest eigenvalues $|\vartheta^{y_{\rm b}}_n|=96.9$, 81.6, and 66.9 [a.u.] from left to right. We can clearly observe that the intensity distribution is enhanced in the region around the brass scatterer (top row), such that removing the scatterer changes the intensity pattern strongly (bottom row).}
\end{figure}

To demonstrate the efficiency of this approach also in the experiment, we implemented a microwave scattering setup as displayed in Fig.~\ref{fig:setup}, consisting of a rectangular multi-mode waveguide made of aluminum with dimensions $L$ $\times$ $W$ $\times$ $H$ = 2.38$~$m $\times$ 10$~$cm $\times$ 8$~$mm (see caption in Fig.~\ref{fig:setup}).
Additional absorbers (types: LS-14 and LS-16 from EMERSON\&CUMING) at the ends of the waveguide mimic semi-infinite leads.
Ten antennas are placed equidistantly in the incident part of the waveguide controlling ten transmission channels individually. The antennas are addressed by IQ-modulators (GTM 1 M2L-68A-5 of GT Microwave Inc.), which allow us to adjust phase and amplitude of the signal emitted by each antenna. The shaping of the incident wave is explained in more detail in \cite{boeh16a}. This array of emitting antennas is connected via a power splitter (Microot MPD16-060180) to a 4 port vector network analyzer (VNA, Agilent E5071C). The transmission through the scattering system is measured with a single movable antenna placed at the output side of the waveguide. In the middle we place 18 cylindrical Teflon scatterers of radius $r=2.55~$mm and one brass scatterer of radius $r_{\rm b}=8.825~$mm forming the disordered scattering system. The relatively high refractive index of the Teflon elements ($n=1.44$) lets all waves undergo multiple scattering events before being transmitted.
We operate the waveguide at 15.5$~$GHz, at which frequency ten propagating transverse electric modes TE$_{0i}$ ($i=1,\dots,10$) are supported.
In the following, the parameter $\alpha$ corresponds to the position $y_{\rm b}$ of the central brass scatterer and the operator $Q_{y_{\rm b}}$ can be computed from the shift $\pm \Delta y_{\rm b}$ from its initial position, which we choose in the experiment to be of the same size as the in-plane radius of the brass scatterer.

\begin{figure}
\centerline{\includegraphics[width=0.94\linewidth]{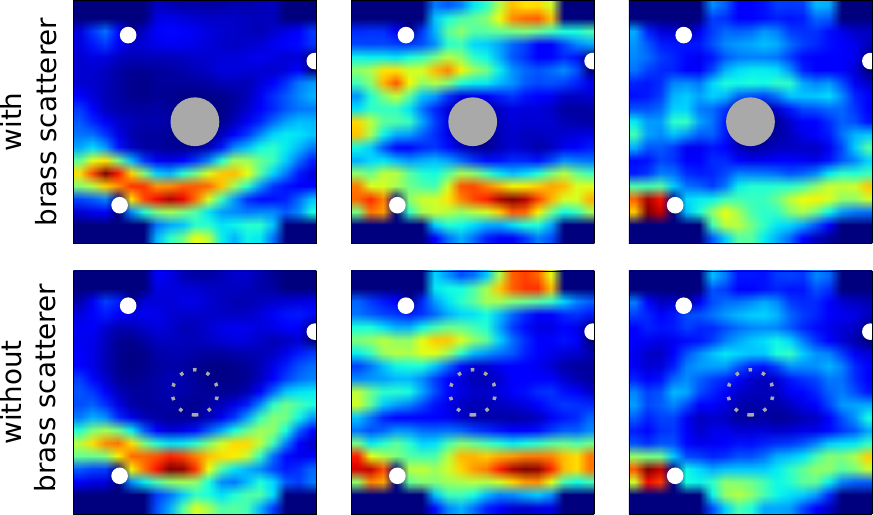}}
\caption{\label{fig:omission}
(Color online.) Same as Fig.~\ref{fig:focus}, but for the eigenstates of $q_{x_{\rm b}}$ with the smallest eigenvalues $|\vartheta^{y_{\rm b}}_n|=1.9$, 2.1, and 6.0 [a.u.] from left to right. The measured intensity pattern clearly avoids the brass scatterer in the middle (top row), such that removing the scatterer leaves the intensity pattern almost unchanged (bottom row).}
\end{figure}

As in most experiments \cite{Popoff2010tm, Yu2013}, we also only have access to a subpart of the entire scattering matrix. Specifically, we can measure only the $10\times10$ transmission matrix $t$, where the complex matrix elements $t_{ji}$ stand for the transmission of the $i$-th antenna of the input antenna array to the $j$-th position of the movable antenna at the output. Even for flux-conserving scattering (without gain or loss), the $t$-matrix is generally non-unitary, since the reflected part of the incident wave is not contained in $t$. The derivation of Eq.~\eqref{eq:static-output} can, however, be easily adapted by replacing $S$ with $t$ and $\omega$ with $y_b$ in Eq.~\eqref{eq:Q-eigenvalue} (see also \cite{Juarez2012}). Most importantly, the resulting non-Hermitian GWS operator,
\begin{equation}\label{eq:d-definition}
q_{y_{\rm b}} := -i t^{-1} \frac{\derd t}{\derd y_{\rm b}},
\end{equation}
inherits the property from its Hermitian counterpart $Q_{y_{\rm b}}$ that its eigenstates are invariant with respect to a small change in the parameter $y_{\rm b}$. In contrast to eigenstates of $Q_{y_{\rm b}}$, the eigenstates of $q_{y_{\rm b}}$ feature injection only from one lead, i.e., $\vec{u}_n \rightarrow \vec{u}_{n,L}$. The transmitted state, i.e., the outgoing state to the right, can be calculated via $\vec{v}_R=t\vec{u}_L$. Specifically, when adapting Eq.~\eqref{eq:static-output} to feature the complex eigenvalues  $\vartheta^{y_{\rm b}}_n=\beta^{y_{\rm b}}_n + i \kappa^{y_{\rm b}}_n$ of $q_{x_{\rm b}}$, we have for the corresponding transmitted states
\begin{equation}\label{eq:static-output-xm}
\vec{v}_{n,R}\!\left(y_{\rm b} + \Delta y_{\rm b} \right) \approx
e^{(i \beta^{y_{\rm b}}_n - \kappa^{y_{\rm b}}_n) \Delta y_{\rm b}} \, \vec{v}_{n,R}\!\left(y_{\rm b} \right)\,.
\end{equation}
Since the construction of the operator $q_{y_{\rm b}}$ involves only the transmission matrix (the reflected part is omitted), its complex eigenvalues  $\vartheta^{y_{\rm b}}_n$ do no longer correspond directly to the local momentum shift $\Delta k^{\rm b}$ at the brass scatterer (see details in the supplemental material \cite{Ambichl2016}). We do find, however, that a strong correlation between these two quantities persists (see Fig.~S3 in \cite{Ambichl2016}), basically since the trace left in the transmission matrix by moving the target scatterer is in itself already quite indicative of the focus on this target. Since this trace in the transmitted signal appears both in the phase and amplitude of a $q_{x_b}$-eigenstate as measured, respectively, by the real ($\beta^{y_{\rm b}}_n$) and imaginary ($\kappa^{y_{\rm b}}_n$) parts of $\vartheta^{y_{\rm b}}_n$, we work with the absolute value of $\vartheta^{y_{\rm b}}_n$ to quantify this trace. A more detailed analysis \cite{Ambichl2016} shows that the correlation between $|\vartheta^{y_{\rm b}}_n|$ and $|\Delta k^{\rm b}|$ can be further increased by normalizing the latter term with the transmission (i.e., by working with $|\Delta k^{\rm b}|/|t^2|$). We can thus apply the concept we derived for the Hermitian GWS operator $Q_{y_{\rm b}}$ also to its non-Hermitian counterpart $q_{y_{\rm b}}$ with the essential difference being that the eigenvalues are now complex and sorted by their absolute value.

Following this protocol also in the experiment, we first measure the transmission amplitudes $t_{ji}$ for two slightly different positions of the brass scatterer ($\pm \Delta y_{\rm b}$). We then determine the GWS operator $q_{y_{\rm b}}$ by replacing the derivative in Eq.~\eqref{eq:d-definition} with a finite-difference approximation based on the difference between the two prior transmission matrix measurements. In a next step we evaluate the eigenstates of $q_{y_{\rm b}}$ and inject them directly through the antenna array at the input port. To test if our focusing protocol works successfully, we then measure the intensity distribution of the microwave field in the vicinity of the brass scatterer by an additional scanning antenna. This antenna is attached to a movable arm and enters through a grid of holes (5$~$mm $\times$ 5$~$mm) in the top plate of the waveguide (extending $3$\,mm into the cavity). The obtained intensity distributions for the eigenvectors with the three largest and the three smallest eigenvalues (in absolute magnitude) are shown in Figs.~\ref{fig:focus} and \ref{fig:omission}, respectively. The displayed intensity profiles demonstrate very clearly that the largest eigenvalues correspond to states focusing on the target, see top row in Fig.~\ref{fig:focus}. The eigenstates with the smallest eigenvalues, in turn, produce intensity patterns which are reduced almost to noise level in the vicinity of the target, see top row in Fig.~\ref{fig:omission}. As an additional test of our protocol, we also recorded the wave intensity patterns after removing the brass scatterer altogether. In case of the focusing states such an intervention drastically changes the overall configuration of the intensity profiles, with intensity maxima near the position of the removed scatterer remaining clearly visible, see bottom row in Fig.~\ref{fig:focus}. In the case of the states that avoid the brass scatterer, the wave intensity pattern remains almost unchanged as a whole when the scatterer is removed, see bottom row in Fig.~\ref{fig:omission}. For the sake of clarity we emphasize here that the measurement of the intensity distribution in the interior of the investigated system serves only for demonstration purposes and is not necessary for implementing our protocol in the first place.

To summarize, we present here an extension of the Wigner-Smith time-delay operator to a whole class of operators with the exciting property of providing eigenstates that focus on or avoid a designated target inside a disordered medium. These generalized Wigner-Smith (GWS) operators require the information stored in a system's scattering matrix, but, as we show here, a construction based on the transmission (or reflection) matrix alone still maintains its useful features. Measuring the transmission matrix of a system has meanwhile not only been demonstrated for the presented case of microwaves, but also for acoustics \cite{Gerardin2014}, seismology \cite{Larose2004}, and recently also for optics \cite{Popoff2010tm, Yu2013}. As a ``guidestar'' \cite{Horstmeyer2015} for focusing  deep in the multiple scattering regime, we use a movable scatterer inside the medium whose spatial shift leaves conspicuous traces in the measured transmission matrix that we exploit for our protocol. In the practical applications that we envision for future implementations, the spatial shift of the target scatterer could, e.g., result from the self-propelled movement of an object inside an otherwise static medium or be excited from the outside through an ultrasound focal spot that can be conveniently scanned through the medium (see \cite{Xu2011, Judkewitz2013} for recent implementations). Since, in particular, many biological media are turbid for optical light but only weakly scattering for ultrasound, such optical focusing techniques enabled by ultrasound have recently already been successfully explored in biomedical imaging (see \cite{Horstmeyer2015} for a review). So far, however, these techniques had to rely on ``optical phase conjugation'' techniques to time-reverse a scattered wave to a scattering target. Our new approach has the advantage that it works without such a time-reversal mirror for light  and that the degree of focusing can be tuned up to the point where a target inside a medium can be entirely avoided rather than focused on. The latter feature may be particularly attractive for imaging techniques for which it is imperative that certain parts of an imaged tissue do not get exposed to radiation. While the presented experiment using ten guided modes serves as a proof-of-principle demonstration, we expect that the full potential of the method can be exploited once many modes are accessible as in the optical domain using spatial light modulators \cite{Mosk2012, rott16}.

\begin{acknowledgments}
P.A., A.B., M.K., and S.R.~are supported by the Austrian Science
Fund (FWF) through project numbers SFB-NextLite F49-P10 and I 1142- N27 (GePartWave).
J.B.~and U.K.~would like to thank the ANR for
funding through Project GePartWave (ANR-12-IS04-0004-01) and the European Commission through
the H2020 programme by the Open Future Emerging
Technology "NEMF21" Project (664828).
\end{acknowledgments}

\section{SUPPLEMENTAL MATERIAL}
\renewcommand\theequation{S\arabic{equation}}		
\setcounter{equation}{0}

\renewcommand\thefigure{S\arabic{figure}}    		
\setcounter{figure}{0} 
\subsection{Definition of the $C_{\alpha}$-Operator}
In the main text, we  defined the GWS-operators $Q_{\alpha}$ using the derivative of the scattering matrix $S(\alpha)$ with respect to the (unspecified) parameter $\alpha$. According to our definition used in Eq.~(3), $C_{\alpha}=-i\derd/\derd\alpha$ is the conjugate operator to the observable $\alpha$. To write this operator as a matrix we expand it in the modes that carry energy from the asymptotic region to the scattering region through its surface and vice versa. We write for the $i$-th mode in the asymptotic region $\psi_n(x, \vec{\xi})$, with $\vec{\xi}=(y,z)^T$ being the transversal coordinate (within the surface) and $x$ the longitudinal coordinate (perpendicular to the surface), respectively. The matrix elements of the operator associated to $C_{\alpha}$  can then be expressed as follows,
\begin{equation}\label{eq:T_alpha}
[C_{\alpha}]_{mn}= \left[ -i\int_{\partial \Omega}^{}d\xi^D \psi^{\star}_m (x,\vec{\xi})\frac{\derd \psi_n(x,\vec{\xi})}{\derd \alpha} \right]_{x = x_s},
\end{equation}
where $\partial \Omega$ denotes the surface with dimension $D$ and $x_s$ is the longitudinal coordinate at which the integral is evaluated.

\subsection{Wave Expectation Values of $Q_{\alpha}$}
In order to obtain a general expression for the operators $Q_{\alpha}$ defined in the main text, we start with the definition of $C_{\alpha}$ in Eq.~\eqref{eq:T_alpha} right above. Using the ``translation'' operator $\exp(iC_{\alpha}\Delta\alpha)$, where $\Delta \alpha$ is the shift in the parameter $\alpha$, we can write for the scattering matrix,
\begin{equation}
S(\alpha + \Delta\alpha) = e^{-iC_{\alpha}\Delta\alpha}S(\alpha)e^{iC_{\alpha}\Delta\alpha}.
\end{equation}
For small deviations $\Delta\alpha$ we can thus approximate,
\begin{align}
\begin{split}
S(\alpha + \Delta\alpha) & \approx (1-iC_{\alpha}\Delta\alpha)S(\alpha)(1+iC_{\alpha}\Delta\alpha) \\
&\approx S(\alpha)-iC_{\alpha}S(\alpha)\Delta\alpha+iS(\alpha)C_{\alpha}\Delta\alpha\,,
\end{split}
\end{align}
where we neglected terms of the order $(\Delta\alpha)^n$ with $n \geq2$. The derivation of $S(\alpha)$ as needed for the construction of $Q_{\alpha}$ can then be expressed as follows,
\begin{equation}
\frac{\derd S}{\derd \alpha}=\lim\limits_{\Delta\alpha \to 0}\frac{S(\alpha + \Delta\alpha)-S(\alpha)}{\Delta\alpha}=iS(\alpha)C_{\alpha} - iC_{\alpha}S(\alpha).
\end{equation}
With $S^{\dagger}S=\mathds{1}$, it follows for $Q_{\alpha}$,
\begin{equation}\label{eq:gws_eval}
Q_{\alpha}=-iS^{\dagger}\frac{\derd S}{\derd \alpha}=C_{\alpha} - S^{\dagger}C_{\alpha}S.
\end{equation}
As discussed in the main text, the GWS-operator $Q_{\alpha}$ measures the shift in $\langle C_{\alpha} \rangle$ the wave experiences due to the scattering process. Note that, strictly speaking, we define here the variable $\alpha$ as a tunable \textit{property of the incident wave} rather than that of the system. This means that, when, e.g., spatially moving the scattering system, as in the case when $\alpha =x$, the parameter $x$ refers to the position of the incident wave relative to the system. Moving the scattering region in the positive $x$-direction while keeping the incoming wavefronts unmoved thus corresponds to a variation of $\Delta x <0$ (in analogy to active/passive transformations). For this case, the expectation values of $Q_x$ read as follows
\begin{equation}
\langle Q_{x} \rangle = \vec{u}^\dagger Q_{x} \vec{u} = \vec{u}^\dagger k_{\rm in} \vec{u} - \vec{v}^\dagger k_{\rm out} \vec{v},
\label{eq:global_gws_expval}
\end{equation}
where we used $C_{x} = k_{\rm in/out}$ in order to emphasize that the operator of the conjugate variable is now the momentum.

For the case where the system consists of $i_{\rm tot}$ scatterers at positions $x_i$ (also the boundaries of the scattering region are considered as individual scatterers), the GWS-operator $Q_{x}$ reads as follows
\begin{equation}
Q_{x}=\sum_{i=1}^{i_{\rm tot}}-iS^{\dagger}\frac{\derd S}{\derd x_i}:=\sum_{i=1}^{i_{\rm tot}} Q_{x_i}=C_{x}-S^{\dagger}C_{x}S,
\end{equation}
where $Q_{x_i}$ stands for the partial GWS-operator corresponding to the positions of the individual scatterers. The sum of all these operators $Q_{x_i}$ is identical to the operator $Q_{x}$, and thus measures the total $\langle C_{x} \rangle$-shift as discussed above [see Eq.~\eqref{eq:gws_eval}]. Hence, each of the summands $Q_{x_i}$ measures the contribution to the total shift in $\langle C_{x} \rangle$, that can be attributed to the respective scatterer at position $x_i$. 

By analogy to Eq.~\eqref{eq:global_gws_expval} we now make the conjecture (to be verified numerically below) that the expectation value of the partial GWS-operator $Q_{x_i}$ can be written as 
\begin{equation}
\langle Q_{x_i} \rangle = \vec{u}^\dagger Q_{x_i} \vec{u} = \vec{u}^{\,i \dagger} k_{\rm in} \vec{u}^{\,i} - \vec{v}^{\,i \dagger} k_{\rm out} \vec{v}^{\,i} = \Delta k^i,
\label{eq:gws_expval}
\end{equation}
where $\vec{u}$ is the incoming wave from the asymptotic regions. The vector $\vec{u}^{\,i}$, on the other hand, represents the local incoming wave directly impinging on the $i$-th scatterer and $\vec{v}^{\,i}=S_i\vec{u}^{\,i}$ is the local wave emanating directly from the considered scatterer. $S_i$ is the scattering matrix describing only the scattering process at the $i$-th scatterer. $k_{\rm in} = -k_{\rm out}$ are the momentum operators in the basis of incoming and outgoing waves. Eq.~\eqref{eq:gws_expval} thus means, that the operator $ Q_{x_i}$ measures the momentum difference $\Delta k^i$ between the local incoming and outgoing waves directly at the shifted scatterer by means of the scattering matrix of the whole system. For $\vec{u}$ being an eigenstate of the partial GWS-operator, the expectation value in Eq.~\eqref{eq:gws_expval} is just the corresponding eigenvalue, i.e., $\langle Q_{x_i} \rangle =\theta^{x_i}$. This operator would thus provide us with a unique tool to extract information about the local scattering process by making use of information that is available in the asymptotic regions. 

\begin{figure}
\includegraphics[width=0.4\textwidth]{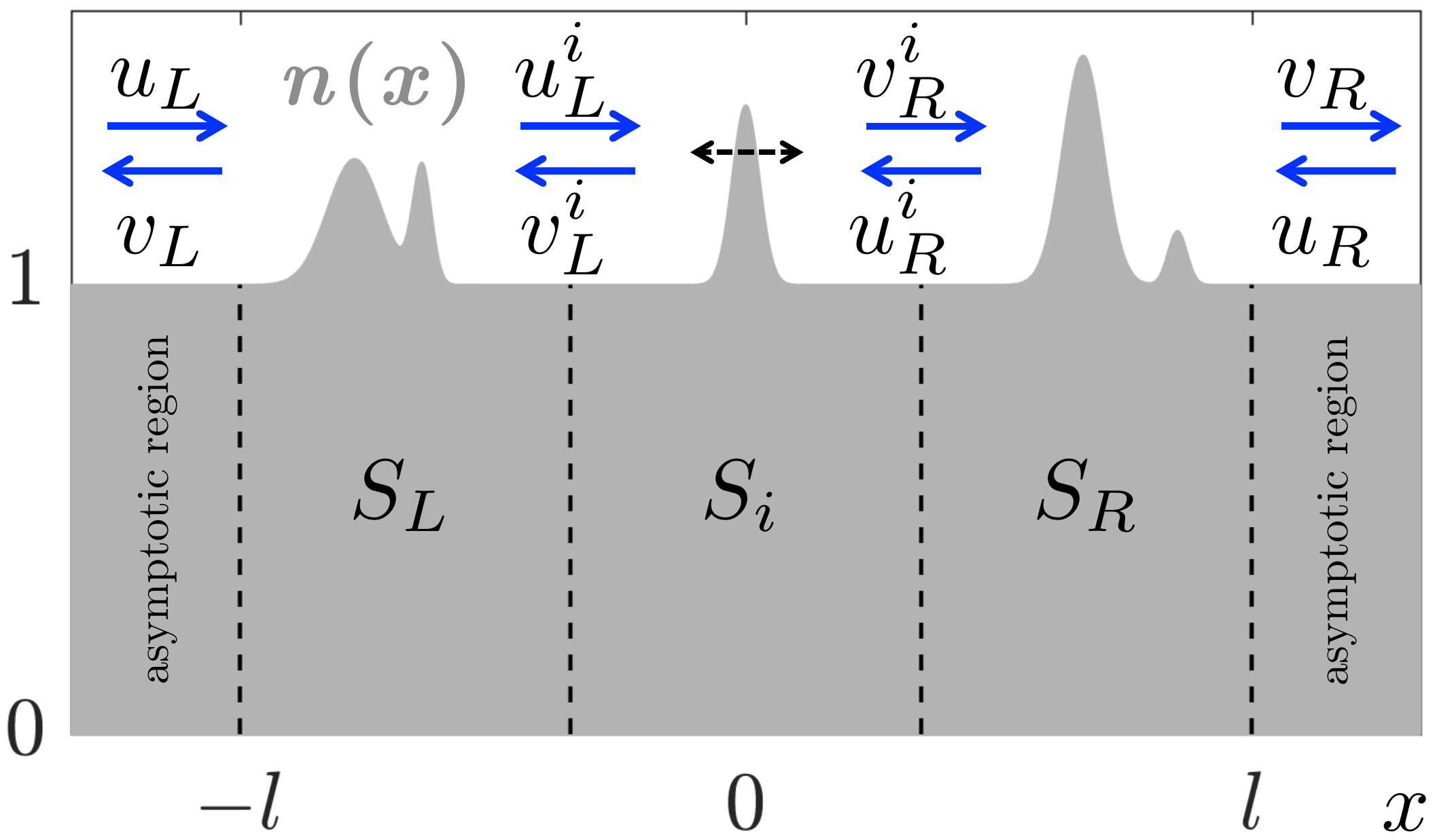}
\caption{(Color online.) Sketch of one possible realization of the considered 1D scattering system located between $-l$ and $l$, where the gray shaded area represents the refractive index $n(x)$. The whole system is divided into three regions: $S_L$ is the scattering matrix of the region on the left-hand side of the shifted scatterer (indicated by the black dashed double-arrow) and $S_R$ is the scattering matrix on the right-hand side of the shifted scatterer. $S_i$ describes the scattering at the shifted scatterer only. $u_{L/R}$ and $v_{L/R}$ are the global incoming and outgoing wave amplitudes from the left/right asymptotic region and the incoming/outgoing amplitudes directly at the shifted scatterer are labeled with the superscript $i$.}
\label{fig:1d_system}
\end{figure}

In order to give an intuitive picture for the above derivation, we consider now a one-dimensional (1D) system in which the momentum operators can be expanded in right- and left-propagating plane waves as 
\begin{equation}
k_{\rm in} = \begin{pmatrix}
k & 0 \\
0 & -k
\end{pmatrix} = -k_{\rm out},
\label{eq:k_ops}
\end{equation}
where $k$ is the scalar wavenumber. Inserting Eq.~\eqref{eq:k_ops} into Eq.~\eqref{eq:gws_expval} and using $\vec{u}^{\,i} = (u^{\,i}_L, u^{\,i}_R)^T$ and $\vec{v}^{\,i} = (v^{\,i}_L, v^{\,i}_R)^T$, we get
\begin{align}
\langle Q_{x_i} \rangle =& \vec{u}^{\dagger} Q_{x_i}\vec{u} \label{eq:gws_weighted} \\ 
=&\underbrace{\Big[\left| u^{\,i}_L \right|^2 k + \left| u^{\,i}_R \right|^2 (-k)\Big]}_{\langle k_{\rm in} \rangle^i} -\underbrace{\Big[ \left| v^{\,i}_R \right|^2 k + \left| v^{\,i}_L \right|^2 (-k) \Big]}_{\langle k_{\rm out} \rangle^i}, \notag
\end{align}
where the subscripts $L$ and $R$ denote the direction from which the waves enter or exit a scattering region. Fig.~\ref{fig:1d_system} shows a generic 1D scattering system consisting of Gaussian scatterers, one of which is the designated target that is shifted to take the derivative $\derd/ \derd x_i$ in $Q_{x_i}$ (see black dashed double-arrow). The global incoming and outgoing waves $\vec{u}$ and $\vec{v}$ as well as the local incoming and outgoing waves directly before and after the scattering process with the shifted scatterer, $\vec{u}^{\,i}$ and $\vec{v}^{\,i}$, are labeled and indicated by blue arrows. Since for the eigenstates the expectation value $\langle Q_{x_i} \rangle$ is just the corresponding eigenvalue $\theta^{x_i}$, Eq.~\eqref{eq:gws_weighted} means that the eigenvalues of such a partial GWS-operator provide us with the information of the momentum shift the local wave acquires directly at the shifted scatterer (weighted with the corresponding intensities [see Eq.~\eqref{eq:gws_expval} and Fig.~\ref{fig:1d_system}].

In order to verify Eq.~\eqref{eq:gws_weighted} [which is the 1D version of Eq.~\eqref{eq:gws_expval}], we now provide an explicit numerical verification based on random matrix theory. For this, we consider the system illustrated in Fig.~\ref{fig:1d_system} consisting of three scattering regions as described by three scattering matrices $S_L$, $S_i$ and $S_R$. $S_{L/R}$ is the scattering matrix of the system to the left/right of the shifted scatterer and $S_i$ is the scattering matrix of the $i$-th scatterer itself (see Fig.~\ref{fig:1d_system}). For the numerical verification we test Eq.~\eqref{eq:gws_weighted} for many random configurations by assuming that $S_L$, $S_i$ and $S_R$ are symmetric and unitary random matrices drawn from the circular unitary ensemble (CUE). For the wavenumber $k$ and the position of the shifted scatterer we use random values. Note, that shifting a scatterer only changes the free propagation length before and after the actual scattering region -- a quantity which we tune explicitly in our numerics. We calculate the eigenvalues $\theta^{x_i}$ of the GWS-operator for each configuration and compare them to the expression on the right-hand side of Eq.~\eqref{eq:gws_weighted}, for which the local amplitudes $\vec{u}^{\,i}$ and $\vec{v}^{\,i}$ are accessible numerically. We find excellent agreement between these two calculated quantities, i.e., the maximal squared absolute error never exceeds $10^{-10}$ for all 1000 system configurations studied -- a deviation which is extremely small compared to the value of $k$ that is of the order of $10^1$.

\begin{figure}
\includegraphics[width=0.45\textwidth]{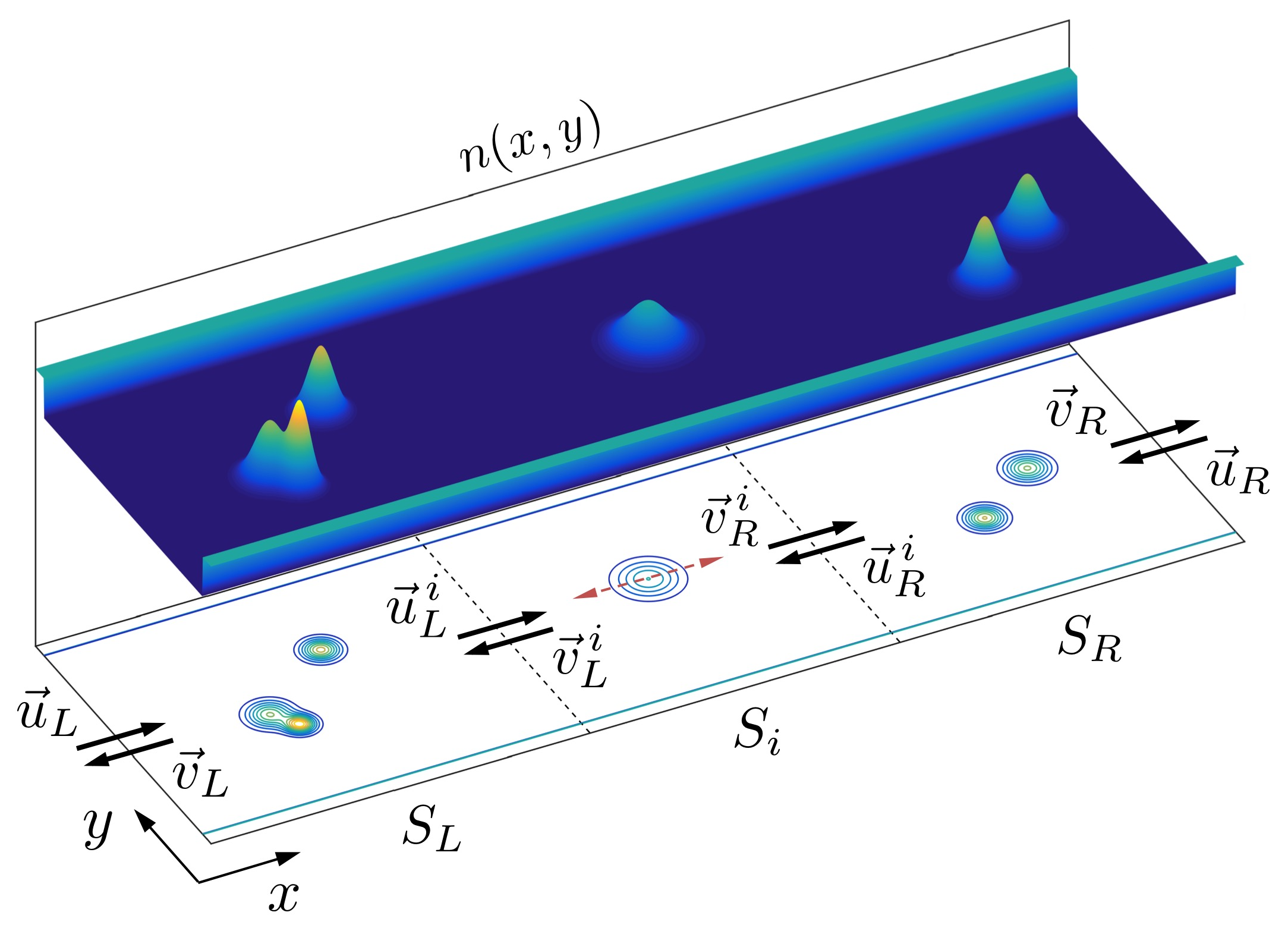}
\caption{(Color online.) Sketch of one possible realization of a considered 2D scattering system, where the top plot shows the refractive index $n(x,y)$ and its corresponding contour lines are projected onto the plane underneath. The whole system is again divided into three regions: $S_L$ is the scattering matrix of the region on the left-hand side of the shifted scatterer (indicated by the red dashed double-arrow) and $S_R$ is the scattering matrix on the right-hand side of the shifted scatterer. $S_i$ describes the scattering at the shifted scatterer only. $\vec{u}_{L/R}$ and $\vec{v}_{L/R}$ are the global incoming and outgoing wave amplitudes from the left/right asymptotic region and the incoming/outgoing amplitudes directly at the shifted scatterer are labeled with the superscript $i$.}
\label{fig:2d_system}
\end{figure}

To show that Eq.~\eqref{eq:gws_expval} is not only valid in 1D, but also in 2D (as in the experiment), we now consider a waveguide structure as shown in Fig.~\ref{fig:2d_system} in which several transverse lead modes can propagate. Also here we test the numerical agreement between the left-hand side of Eq.~\eqref{eq:gws_expval} with its right-hand side for many random configurations, in analogy to the 1D procedure described above.  The difference in 2D is, however, that all scattering matrices have dimension $2N\times2N$, where $N$ is the number of propagating modes. In our simulations we take $S_L$, $S_R$ from th CUE, but the scattering matrix for the middle part, $S_i$, is obtained by an explicit numerical calculation for a single scatterer randomly placed in a 2D waveguide \cite{Rotter2000, Libisch2012}. We calculate $Q_{x_i}$ through shifting this scatterer (in propagation direction) for each random configuration. In this way we obtain excellent agreement between the $Q_{x_i}$-eigenvalues [left-hand side of Eq.~\eqref{eq:gws_expval}] and the local momentum shift [right-hand side of Eq.~\eqref{eq:gws_expval}], which are independently calculated with our numerics. As in 1D, we find that the maximal squared absolute error stays below $10^{-10}$ for all system configurations, which is again small compared to the mode wavevectors that are of the order of $10^1$. This verifies our assumption that the GWS-operator connects asymptotic quantities stored in the scattering matrix of the system with local wave amplitudes at the shifted scatterer. For reasons of simplicity, we shifted the scatteres in the 1D and 2D verification in propagation direction. However, in the derivation of $Q_{\alpha}$ in Eq.~\eqref{eq:gws_eval}, no direction is distinguished from another one, such that Eq.~\eqref{eq:gws_expval} also holds for shifting the scatterer in any other direction (such as the transverse direction considered in the experiment).

\subsection{Wave Expectation Values of $q_{x_i}$}
As in our experiment, where we only have access to a subpart of the full scattering matrix $S$, we use the non-Hermitian GWS-operator (4) based on the system's transmission matrix $t$ in order to find states that focus on or omit a designated scatterer. Eigenstates of both, the Hermitian GWS-operator $Q_{x_i}$ as well as the non-Hermitian operator $q_{x_i}$ used in the experiment are to first order invariant with respect to a change of the position of the chosen scatterer (up to a global factor). Note that eigenstates of $Q_{x_i}$ generally require injection from both leads, whereas eigenstates of $q_{x_i}$ are injected only from one lead. The eigenvalues of $Q_{x_i}$ are real and correspond to the local momentum shift as shown above. The eigenvalues of $q_{x_i}$, however, are complex but can still be related to the local momentum shift as we will show in the following: Writing the Hermitian GWS-operator $Q_{x_i}$ in its block structure
\begin{align}
\begingroup
\renewcommand*{\arraystretch}{1.25}
Q_{x_i}=
	 \begin{pmatrix}
		Q_{x_i}^{11} & Q_{x_i}^{12}  \\
		Q_{x_i}^{21} & Q_{x_i}^{22}
	\end{pmatrix}\,,
\endgroup
\end{align}
we consider now the upper left $Q_{x_i}^{11}$-block, which takes the explicit form $Q_{x_i}^{11}=-i(t^{\dagger}\frac{\derd t}{\derd x_i} + r^{\dagger}\frac{\derd r}{\derd x_i})$, with $t$ and $r$ being the transmission and reflection matrix, respectively. Calculating the expectation value of $Q_{x_i}$ for a wave entering only from the left lead, i.e., $\vec{u}=(\vec{u}_L, 0)^T$, we end up with $\langle Q_{x_i} \rangle =  \vec{u}^\dagger Q_{x_i} \vec{u}=  \vec{u}_L^\dagger Q_{x_i}^{11} \vec{u}_L =\langle Q_{x_i}^{11} \rangle= \langle k_{\rm in} \rangle^i - \langle k_{\rm out} \rangle^i = \Delta k^i$, where $\Delta k^i$ is the local momentum difference. Using the expression for $q_{x_i} = -i t^{-1} \frac{\derd t}{\derd x_i}$, we can rewrite $Q_{x_i}^{11}=-i(t^{\dagger}\frac{\derd t}{\derd x_i} + r^{\dagger}\frac{\derd r}{\derd x_i})= t^{\dagger}t q_{x_i} -ir^{\dagger}\frac{\derd r}{\derd x_i}$. Calculating now the expectation value of $Q_{x_i}^{11}$ for an eigenstate of $q_{x_i}$ (in the following we will omit the eigenstate index for the sake of readability) and resolving for the complex eigenvalue $\vartheta^{x_i}$ of $q_{x_i}$ yields
\begin{equation}
\vartheta^{x_i} = \frac{1}{|t|^2} \left( \Delta k^i - \left\langle ir^{\dagger}\frac{\derd r}{\derd x_i} \right\rangle \right) ,
\label{eq:tau_abmk}
\end{equation}
where $|t|^2 = \langle t^\dagger t \rangle$ is the global transmittance of the eigenstate. As in the case of the Wigner-Smith time-delay operator, one can show that the eigenvalue can also be written in the following form \cite{AmbichlPhd2016}
\begin{equation}
\vartheta^{x_i} = \frac{\derd \varphi_t}{\derd x_i} - i \frac{\derd \ln \left( |t \vec{u}_L| \right)}{\derd x_i} \ ,
\label{eq:tau_phil}
\end{equation}
where the real part is the derivative of the global transmitted scattering phase $\varphi_t$ and the imaginary part can be interpreted as the change in the global output intensity with respect to a change of the local position of the $i$-th scatterer. It is worth noting that this relation still holds for a variation of a local parameter rather than a global one as in the case of time-delay. The reason for this is that in both the global and the local case the output vector is invariant to first order with respect to a change in a certain parameter (e.g., $\omega$ or $x_i$). Comparing Eq.~\eqref{eq:tau_abmk} with \eqref{eq:tau_phil} now shows that the imaginary part of $\vartheta^{x_i}$ can be related to (global) reflections, i.e.,
\begin{equation}
\mathrm{Im} (\vartheta^{x_i}) = - \frac{\derd \ln \left( |t \vec{u}_L| \right)}{\derd x_i} = - \mathrm{Im} \left( \frac{1}{|t|^2} \left\langle ir^{\dagger}\frac{\derd r}{\derd x_i} \right\rangle \right),
\label{eq:tau_imag}
\end{equation}
which have been omitted in the construction of $q_{x_i}$. However, the reflection term in Eq.~\eqref{eq:tau_abmk} also features a real part which means that the real part of $\vartheta^{x_i}$ is not just given by $\Delta k^i/|t|^2$. 

\begin{figure}
\includegraphics[width=0.35\textwidth]{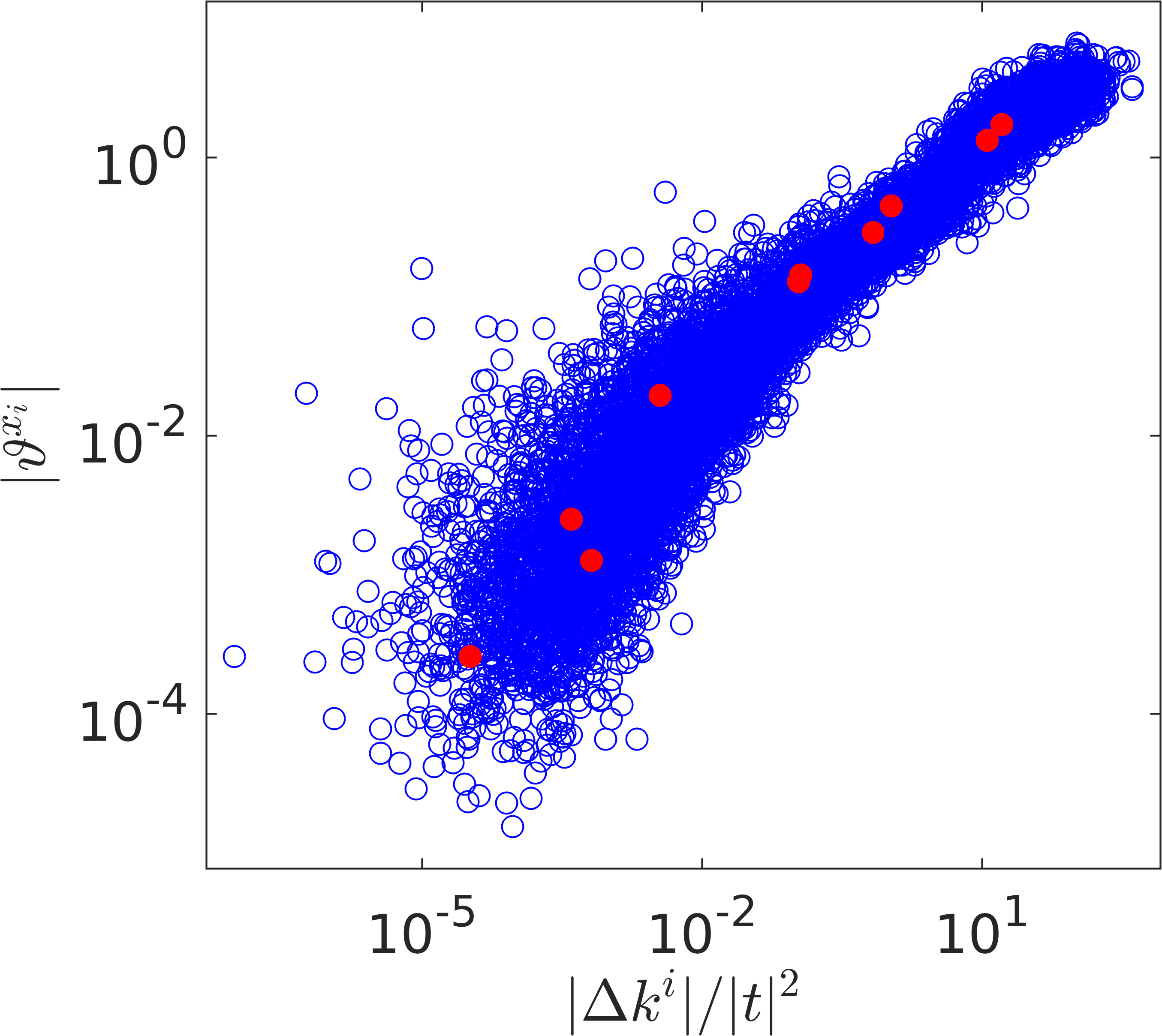}
\caption{(Color online.) Plot of the absolute value of $\vartheta^{x_i}$ versus $|\Delta k^i|/|t|^2$ for a waveguide with 10 modes. The blue circles show the distribution for all 1000 disorder configurations and the red dots correspond to a typical distribution of a single configuration. A correlation between $|\vartheta^{x_i}|$ and $|\Delta k^i|/|t|^2$ can clearly be observed.}
\label{fig:theta_dkt2}
\end{figure}

To assign further meaning to the real part of $\vartheta^{x_i}$, we write, in analogy to the time-delay operator, the expectation value of $Q_{x_i}$ as a sum of derivatives of the scattering phases of each mode weighted with its corresponding output intensity \cite{AmbichlPhd2016} and split the sum into two parts. One part covers the reflected waves and the other one covers the transmitted waves, respectively:
\begin{align}
\langle Q_{x_i} \rangle & = \sum_{n=1}^{2N} \left| (S\vec{u})_n \right|^2 \frac{\derd \varphi_n}{\derd x_i} \label{eq:Q_sum} \\
&= \sum_{n=1}^N \left| (r\vec{u}_L)_n \right|^2 \frac{\derd \varphi_{r,n}}{\derd x_i} + \sum_{n=1}^N \left| (t\vec{u}_L)_n \right|^2 \frac{\derd \varphi_{t,n}}{\derd x_i} , \notag
\end{align}
where $N$ is the number of modes. For the evaluation of this expectation value with a $q_{x_i}$-eigenstate, we use the fact that the transmitted phase derivatives are the same for all modes, i.e., $\frac{\derd \varphi_{t,n}}{\derd x_i} = \frac{\derd \varphi_t}{\derd x_i} = \mathrm{Re}(\vartheta^{x_i})$. Thus, bringing this factor in front of the sum, using $\langle Q_{x_i} \rangle = \Delta k^i$ and resolving for the real part of $\vartheta^{x_i}$ gives
\begin{equation}
\mathrm{Re}(\vartheta^{x_i}) = \frac{\derd \varphi_t}{\derd x_i} =  \frac{1}{|t|^2} \left( \Delta k^i - \sum_{n=1}^N \left| (r\vec{u}_L)_n \right|^2 \frac{\derd \varphi_{r,n}}{\derd x_i} \right) .
\label{eq:tau_real}
\end{equation}
Eq.~\eqref{eq:tau_real} shows that the real part of the $q_{x_i}$-eigenvalue, $\mathrm{Re}(\vartheta^{x_i})$, contains both the momentum shift $\Delta k^i$ as well as the global reflectivity of each mode multiplied by the corresponding phase derivatives. Thus, the desired correlation between a small/large $|\mathrm{Re}(\vartheta^{x_i})|$ (or $|\vartheta^{x_i}|$) and a small/large momentum shift $|\Delta k^i|$ is not as obvious as for the eigenvalues of $Q_{x_i}$. We therefore investigate with our numerical approach (see above), which quantity based on $\vartheta^{x_i}$ shows the strongest possible correlation with $\Delta k^i$. Testing several different expressions, we find the best results for $|\vartheta^{x_i}|$ and $|\Delta k^i|/|t|^2$ (see Fig.~\ref{fig:theta_dkt2}). Specifically, the probability for finding the two states (out of ten) with the highest/smallest values of $|\Delta k^i|/|t|^2$ among the states with the three highest/smallest $|\vartheta^{x_i}|$ is 99.9\%/91.7\% in our 2D random matrix model featuring 10 modes and 1000 random configurations. This observation allows us to relate the eigenvalues of $q_{x_i}$-eigenstates with its focusing strength even without knowledge of the reflection matrix. As we show in the next section, it is also possible to go beyond such a semi-empirical analysis by applying a special projection procedure to our $q_{x_i}$-operator, which restores a strong correlation between $|\mathrm{Re}(\vartheta^{x_i})|$ and $|\Delta k^i|$ itself.

\subsection{Wave Expectation Values of $q_{x_i}$ for Singular Transmission Matrices $t$}
Since the construction of the non-Hermitian operator $q_{x_i}$ involves the inverse of the transmission matrix $t$, we have to deal with the problem that $t$ can become singular. This occurs, e.g., when transmission channels are closed. As a result, a straightforward inversion of the transmission matrix fails, requiring instead the evaluation of an effective inverse by projecting the transmission matrix onto a subspace of highly transmitting channels. For the projection procedure we make use of a singular value decomposition (SVD) of the transmission matrix $t = U \Sigma V^\dagger$, where the matrices $U$ and $V$ contain column-wise the left and right singular vectors and the matrix $\Sigma = \mathrm{diag}(\{\sigma_n\})$ contains the singular values $\sigma_n$ on its diagonal. In a next step, we pick a certain subset of large singular values $\tilde{\Sigma} = \mathrm{diag}(\{ \tilde{\sigma}_n \})$ and corresponding left and right singular vectors $\tilde{U}$ and $\tilde{V}$. With these matrices we can construct an effective inverse $t^{-1} = \tilde{V} (\tilde{U}^\dagger t \tilde{V})^{-1} \tilde{U}^\dagger = \tilde{V} \tilde{\Sigma}^{-1} \tilde{U}^\dagger$, where $\tilde{\Sigma}^{-1} = \mathrm{diag}(\{ \tilde{\sigma}_n^{-1} \} )$. 
Projecting also the derivative of $t$ onto this subspace, using the proper projection operators $P_{\tilde{U}} = \tilde{U} \tilde{U}^\dagger$ and $P_{\tilde{V}} = \tilde{V} \tilde{V}^\dagger$, we arrive at the following generalized construction rule for our operator
\begin{equation}
\tilde{q}_{x_i} = -i \ \tilde{V} (\tilde{U}^\dagger t \tilde{V})^{-1} \tilde{U}^\dagger \ \tilde{U} \tilde{U}^\dagger \frac{\derd t}{\derd x_i} \tilde{V} \tilde{V}^\dagger
\label{eq:q_projection}
\end{equation}
which turns into our original expression for $q_{x_i}$ if all transmission channels, i.e., all singular values, are taken into account.

Apart from applying this procedure to non-invertible transmission matrices, it can also be used to restore a correlation between $|\mathrm{Re}(\tilde{\vartheta}^{x_i})|$ and $|\Delta k^i|$ since the projection on highly transmitting channels suppresses the reflection term in Eq.~\eqref{eq:tau_real} and increases the global transmittance of $\tilde{q}_{x_i}$-eigenstates. In order to numerically verify this, we again consider many random 2D configurations as above and look at the statistical correlation between these two quantities with and without cutting away reflecting channels. To quantify the strength of this correlation, we look at the probability for finding the state with the highest/smallest $|\Delta k^i|$ among the states with the two highest/smallest $|\mathrm{Re}(\tilde{\vartheta}^{x_i})|$. Without SVD these probabilities are 63.7\%/79.3\%, whereas the application of our procedure using the five highest transmitting states (out of ten) yields 95.5\%/88.3\%. This verifies that projecting onto highly transmitting channels provides the possibility to restore a correlation between $|\mathrm{Re}(\tilde{\vartheta}^{x_i})|$ and $|\Delta k^i|$.

\subsection{Data Processing}
For the processing of the data measured in the experiment we Fourier filter the transmission signal between the antenna array and the scanning antenna. The resulting intensity data are treated with a discrete two dimensional convolution of the form,
\begin{equation*}
c_{s}(x,y)=s*K (x,y) = \sum_{x^{\prime},y^{\prime}} K(x^{\prime},y^{\prime}) s(x-x^{\prime},y-y^{\prime})\,,
\label{eq:corrletaionfuncdisc}
\end{equation*}
with
\begin{equation*}
K = \frac{1}{4}
	 \begin{pmatrix}
		0.25 & 0.5  &  0.25\\
		0.5 & 1 & 0.5\\
		0.25 & 0.5  &  0.25
	\end{pmatrix}\,,
\end{equation*}
to generate Fig.~2 and 3. The parameter $s$ denotes the intensity of the measured transmission signal and $K$ is called the convolution kernel, which in our case takes only closely neighboring measurements (pixels) into account. Note that $x,~y,~x^{\prime}$ and $y^{\prime}$ are discrete quantities corresponding to the coordinates of the grid-holes of the scanning antenna.

\end{document}